\documentclass[useAMS,usenatbib]{mn2e}
\usepackage{graphicx}
\usepackage{psfig}
\usepackage{amsmath}
\usepackage{amsfonts}
\usepackage{amssymb}
\usepackage{color}
\usepackage{times}

\newcommand{\dl}{\delta}

\newcommand{\sg}{\sigma}
\newcommand{\Nc}{\mathcal{N}}

\newcommand{\rbh}{\bar{\rho}}

\def\msun{\mbox{$M_{\odot}$}}

\title[Mass function and assembly histories of dark halos]
{Mass function and assembly of dark haloes: an approach to inventory
isolated overdense regions in random fields}

\author[Firmani \& Avila-Reese]
{C. Firmani$^{1,2}$\thanks{E--mail: firmani@merate.mi.astro.it}, 
V. Avila-Reese$^{1}$\thanks{E--mail: avila@astro.unam.mx}\\
$^{1}$Instituto de Astronom\'{\i}a, Universidad Nacional Aut\'onoma de M\'exico,
A.P. 70-264, 04510, M\'exico, D.F.\\
$^{2}$Osservatorio Astronomico di Brera, via E.Bianchi 46, I-23807
Merate, Italy}

\begin{document}

\pagerange{\pageref{firstpage}--\pageref{lastpage}} \pubyear{2002}
\maketitle
\label{firstpage}

\begin{abstract}

In order to attain a statistical description of the evolution of cosmic density fluctuations in 
agreement with results from the numerical simulations, we introduce a probability conditional formalism (CF) 
based on a complete inventory of isolated overdense regions in a density random field.
This formalism is a useful tool for describing at the same time the mass function (MF) of 
virialized dark haloes, their mass aggregation histories (MAHs) and merging rates (MRs).
The CF focuses on virialized regions in a self-consistent way rather than in mass elements, 
and it offers an economical description for a variety of random fields. 
Within the framework of the CF, we confirm that, for  a Gaussian field, it is not possible
to reproduce at the same time the MF, MAH, and MR of haloes, both for a constant and moving barrier.
Then, we develop an inductive method for constraining the cumulative conditional
probability from a given halo MF description, and thus, using the CF, we calculate 
the halo MAHs and MRs.
By applying this method to the MF measured in numerical simulations by Tinker et al., we find 
that a reasonable solution, justified by a mass conservation argument, is obtained if a rescaling 
--increment by $\sim 30\%$-- of the virial mass defined in simulations is introduced, and
a (slight) deviation from Gaussianity is taken into account. 
Thus, both the MAH and MR obtained by a Monte Carlo merger tree agree now with the 
predictions of numerical simulations.
We discuss on the necessity of rescaling the virial mass in simulations when comparing
with analytical approaches on the ground of the matter not accounted as part of the halos 
and the halo mass limit due to numerical resolutions in the simulations.
Our analysis supports the presence of a diffuse dark matter component that is not taken 
into account in the measured halo MFs inasmuch as it is not part of the collapsed structures.

\end{abstract}


\section{Introduction}

According  to the contemporary cosmological paradigm, cosmic structures emerge from the 
gravitational growth of primordial dark matter density fluctuations.  
A central problem in the last four decades has been the connection between this primordial fluctuation 
field and the abundance and assembly history of the virialized dark matter haloes; in more detail,
their mass function (MF), mass aggregation histories (MAHs), and merger rates (MRs).
Since the seminal work by Press \& Schechter (1974; hereafter PS), many statistically-based analytical 
formalisms were developed in order to perform such a connection (see for reviews e.g., Zentner 2007; 
Mo, van den Bosch \&  White 2010). 

In order to achieve an association between the density fluctuation field and the virialized haloes, one needs: 
a statistical description of the density fluctuation field, an inventory of the overdense regions which 
will be associated to the virialized objects, and a model for the dynamical evolution of the overdensities, 
including an operational criterion of collapse (virialization). 

PS assumed a Gaussian density field and used the spherical collapse model (Gunn \& Gott 1972). 
A region of the density field is assumed to end as a virialized halo when its linearly evolving overdensity 
exceeds a critical value $\dl_c$. The virialized mass at a fixed scale is assumed to be given
by the contributions of the regions of this scale overdense by $\dl_c$ (the \textit{PS Ansatz}) 
plus the underdense regions of the same scale contained in larger regions overdense by $\dl_c$; 
the latter statement enunciates the so-called {\it cloud-in-cloud problem}. In order to account for all 
mass, PS assumed that the second contribution equals to the first one, which justifies this the fudge 
factor of 2 introduced in their inferred halo MF. 
 
Based on a more rigorous statistical description, the excursion set (ES), also know as extended PS (EPS) 
formalism (Peacock \& Heavens 1990; Bond et al. 1991), allows to overcome the cloud-in-cloud problem.
The ES formalism applied to a Gaussian field provides a tool to compute, besides the unconditional MF, 
the conditional MF that can be used for generating halo merger trees (Bond et al. 1991; Lacey \& Cole 1993). 
In this formalism, the result depends critically on the filter used to average the linear density fluctuation 
field at different mass scales.  
The mathematical solution is straightforward when a sharp $k$-space filter is adopted. 
For this filter, the PS result including the fudge factor of 2 is recovered. 
In the calculation of the merger trees, the  sharp $k$-space filter implies independent steps (Markovian random 
walks) along the mass trajectory.
Note that the ES formalism has a conceptual problem for predicting the halo mass in which a particular mass 
element ends up (Mo et al. 2010, \S\S 7.2.2), problem that may lead to an inaccurate buildup of MAHs and MRs. 

With the advent of large cosmological N-body simulations, the whole non-linear process of dark matter 
gravitational evolution and collapse into virialized haloes could be followed, though with strong limitations 
due to mass resolution. 
Do approaches based on the Gaussian ES formalism allow to describe correctly the results from the simulations? 
In particular, {\it do results from these approaches agree at the same time with the MF, MAHs, and MRs of haloes 
as measured in simulations?}

A non-negligible discrepancy between the MFs obtained in the ES formalism and the N-body simulations has been 
early reported. 
The introduction in the ES formalism of the elliptical gravitational collapse instead of the spherical one  
(mass-dependent instead of constant $\dl_c$, respectively) helped to overcome this problem (Sheth, Mo \& 
Tormen 2001).
However, in this case the ES formalism does not provide an analytic formula for the conditional MF, and the 
merger trees based on the Gaussian ES formalism show deviations with respect to simulations in the progenitor 
mass distributions (Sheth \& Tormen 2002), the mass contained within all the progenitors (Neistein et al. 2006), 
and the average main progenitor MAHs and MRs (Wechsler et al. 2002; van den Bosch 2002).
On the other hand, direct measures of the conditional MFs in N-body simulations show that they depart 
from the corresponding functions calculated with the ES formalism for a Gaussian field (Cole et al. 2008; 
Neistein et al. 2010). 
Thus, a possible source of the discrepancies lies in the strict assumption of Gaussianity in the ES formalism. 
Note that the question is not about primordial non-Gaussianity; the N-body simulations use indeed a Gaussian 
density field that is evolved analytically (e.g., by means of the Zel'dovich approximation) until the quasi-linear 
regime is reached. 
The point is that non-negligible deviations from Gaussianity at the scales of interest have likely happen already 
during this regime (Coles \& Jones 1991).
Here, we set aside Gaussianity and the ES formalism and handle the problem of the inventory of overdense regions 
in arbitrary random fields. 

We present an approach aimed to state the problem of the inventory of overdense regions in random 
(Gaussian or non-Gaussian) fields from a very general point of view. 
We assume that the density fluctuation field is fully characterized by its field conditional probability function. 
In order to face the problem of associating the halo mass to the overdensity in an alternative 
way to the ES formalism (see the reference Mo et al. 2010 cited above) we proceed as follows.
In our {\it conditional formalism} (hereafer CF), we make use of  the concept of {\it isolated regions} introduced by 
Jedamzik (1995; see also Yano, Nagashima \& Gouda 1996; Nagashima 2001) as a natural way to connect 
overdensities to virialized regions. 
We enunciate the {\it isolated overdense regions inventory} theorem, through which the given field conditional 
probability function of the random field is linked to the inventory of isolated overdense regions inside larger 
isolated regions of lower overdensity, being the former those that eventually collapse into virialized haloes if 
their overdensities equal to $\dl_c$.

The CF offers an alternative EPS formulation.  
It gives an easy way to develop a Monte Carlo algorithm for calculating the MAH and MR of the growing virialized 
haloes, because it supplies naturally the progenitor conditional probability function
\footnote{For brevity, hereafter we will omit the specification "progenitor" when referring to the conditional 
probability.}
of finding a specific isolated region (progenitor) inside a given isolated region (descendent).
When the problem of finding isolated overdense regions inside larger ones is extended to 
the overall Universe, then such conditional probability reduces to an unconditional probability,
which gives the halo MF.
Nevertheless, if one starts from an unconditional probability function, for example on the basis of a 
halo MF obtained from N-body simulations, and tries to go back to the conditional probability function, 
in order to build merger trees compatible with the given MF, then multiple possibilities appear.
The CF offers useful tools to make a choice. 
As we will discuss in \S\S 4.2, we decide to leave to the Gaussian hypothesis even if, for simplicity, 
we retain the (Gaussian) PS Ansatz in a generalized version.
 
The CF might be a powerful tool for economically describing {\it in a consistent way the simulation 
results related to the MF, MAHs, and MRs of the virialized haloes}. 
These descriptions can then be easily implemented in semi-analytical and semi-empirical models of galaxy 
evolution, and can be extended to masses and epochs, where the resolution is a limit for the simulations. 
On the other hand, the CF may allow to explore economically cosmic structure formation in alternative 
cosmologies or in cases where the density fluctuation field is intrinsically non-Gaussian.
Our approach could be applied also in other astrophysical problems, where random density fields
are introduced, e.g., the formation and evolution of dense gas structures in the interstellar
medium as giant molecular clouds, massive clumps and cores (Hopkins 2012). 

This paper is organized as follows. Section 2 is devoted to present the CF;
in \S\S 2.1, an equation for the isolated overdense regions inventory is derived, and
in \S\S 2.2 the algorithms to build up the merger trees are introduced.
Section 3 deals with the particular case of Gaussian random fields. 
The halo MF is derived in \S\S 3.1, while the MAH and MR are calculated in \S\S 3.2. 
A discussion on the moving barrier effects is presented in \S\S 3.3. 
In view of the discrepancies obtained when using as input the Gaussian statistics,
in Section 4 an heuristic approach, based on the use of the halo MF from 
simulations as input, is presented. 
In \S\S 4.1 we approach the problem of the difference on the mass estimates in the 
analytic method and in the simulations, and remark the necessity to introduce
a mass rescaling,  representative of a diffuse matter component, to connect the two 
standpoints. The results concerning MAHs and MRs are presented in \S\S 4.2.
Our conclusions are given in Section 5.
We adopt a flat $\Lambda$CDM cosmology with $\Omega_M= 0.27, \Omega_\Lambda=0.73,
h=0.71, \sigma_8=0.9, n=1$.

\section{The conditional formalism}

The ES formalism provides a means to derive the halo MF, accounting for the cloud-in-cloud problem, 
and to build up the halo MAHs. 
However, as mentioned above, the Gaussian ES formalism seems to have difficulties for predicting at the 
same time the MF, MAH, and MRs in agreement with numerical simulations.
Also, with the ES formalisms it is not easy to deal with non-Gaussian random fields (Inoue \& Nagashima 2002),
though some progress has been made recently by introducing a stochastic barrier and non-Markovian
corrections (Maggiore \& Riotto 2010), by manipulating the step-size distribution of the random walks (Lam \& Sheth 2009)
or by accounting for correlations between steps (Musso \& Paranjape 2012).
On the other hand, being the ES formalism focussed on a generic mass element, the virialized halo mass is
associated to a smoothing scale, which may lack of a real physical meaning.
In order to overcome these shortcomings, we introduce below the conditional formalism (CF) inspired by 
the Jedamzik (1995) approach. 
This formalism is based on a complete inventory of the isolated overdense regions of a random density field 
described by a {\it cumulative field conditional probability function}.

\subsection{A complete inventory of isolated overdense regions}

Let start with some key definitions.

\noindent 
{\it Isolated regions}:
an isolated region with overdensity $\dl$ is a connected region not included in a larger connected region 
with overdensity $\geq \dl$, while at the same time, outside of the totality of isolated regions with 
overdensity $\dl$ there does not exist any region with overdensity $\geq \dl$. 
Hereafter we will refer to an isolated region with overdensity $\dl$ by means of the symbol ${\bf I}(\dl)$  
\footnote{
The definition of isolated region allows to introduce new mathematical concepts.
Given a space with a distribution of isolated regions ${\bf I_n}(\dl)$, one can cut an arbitrary 
region of such space along a specific border.
The consequence will be that some ${\bf I_n}(\dl)$ will be internal, others will be external and 
finally some of them will be on the border of that region.
By definition, when no one of the isolated regions ${\bf I_n}(\dl)$ will be on the border, such region  
will be named {\it suitable} with respect to its internal isolated regions overdense by $\dl$.
It is straightforward to see that an isolated region ${\bf I}(\dl)$ is a {\it suitable} region with
respect to all its internal isolated regions more overdense than $\dl$.
Another example is given by an isolated region ${\bf I}(\dl)$ where one (or more) internal isolated 
region(s) ${\bf I_r}(\dl')$ with $\dl' > \dl$ has (have) been removed.
The result is a {\it suitable} region $\bf S$ with respect to its internal isolated regions 
${\bf I}(\dl')$, its overdensity is less than $\dl$, and typically it will not be an isolated region.
Any region of $\bf S$ more overdense than $\dl'$ is contained inside an internal isolated region 
${\bf I}(\dl')$.
Afterwards when we will refer to a region $\bf S$ actually we will refer to a region {\it suitable} 
with respect to the internal isolated regions overdense by $\dl$ involved by the problem.}.

\noindent
{\it Virialized regions}:
a virialized region is an isolated region whose overdensity $\dl$ is equal to a critical value $\dl_c$.

\noindent
$\phi( M' , \dl' \vert M , \dl) d\dl'$:
is the  {\it field conditional probability} to find a region of mass $M'$ with an overdensity between $\dl'$
and $\dl'+d\dl'$, contained inside a larger region of mass $M$ and overdensity $\dl$.

\noindent
$\Nc( M' , \dl' \vert M , \dl) dM'$:
is the number of isolated regions, i.e. the {\it conditional MF}, with overdensity $\dl'$ and
mass between $M'$ and $M'+dM'$, contained inside a larger region $\bf S$ with overdensity $\dl$ and mass $M$.

\noindent
The {\it cumulative field conditional probability} to find a region of mass $M'$ with overdensity $\geqslant \dl'$ 
inside a larger region of mass $M$ and overdensity $\dl$ is
\begin{equation}
F( M',\dl' \vert M,\dl) = \int_{\dl'}^{\infty}~\phi( M' , \dl'' \vert M , \dl) d\dl''
\nonumber
\end{equation}

\noindent
Our CF concerns random fields that can be defined by such conditional probability.

Following, a theorem focussed on an {\it isolated overdense regions inventory} and aimed to establish a 
connection between the cumulative field conditional probability, 
$F( M',\dl' \vert M,\dl)$, and the conditional MF,
$\Nc( M' , \dl' \vert M , \dl)$ with $\dl' \geq \dl$, is presented.

Consider a region $\bf S$ with overdensity $\dl$ and mass $M$.
Define an arbitrary mass $M'_1 \leqslant M$.
A decomposition of the mass range $(M'_1, M)$ into $n$ ordered steps, $dM'_i$, identifies a sequence 
of $d\Nc_i = \Nc(M'_i , \dl' \vert M , \dl) dM'_i$ isolated regions ${\bf I}(\dl')$ with overdensity 
$\dl'$ and masses between $M'_i$ and $M'_i+dM'_i$ inside $\bf S$.

For any given $\dl'' \geqslant \dl'$, the amount of mass in $\bf S$ assembled by regions with 
mass $M'_1$ and overdensity between $\dl''$ and $\dl''+d\dl''$ is
\begin{equation}
M \phi( M'_1,\dl'' \vert M,\dl) d\dl''
\nonumber
\end{equation}
This mass is also given by the sum of the regions of mass $M'_1$ with overdensity between 
$\dl''$ and $\dl''+d\dl''$ located in each one of the isolated regions ${\bf I}(\dl')$ with masses 
between $M'_i$ and $M'_i+dM'_i$, and overdensity $\dl'$ of the above defined sequence:
\begin{equation}
\sum_{i=1}^n M'_i \phi( M'_1,\dl'' \vert M'_i,\dl') d\dl''\ d\Nc_i\  
\nonumber
\end{equation} 
No other contribution exists because outside the totality of the isolated regions ${\bf I}(\dl')$ 
in $\bf S$ there are not regions with overdensity $\geq \dl'$.
By equating the last two contributions, we can write
\begin{equation}
\phi( M'_1,\dl'' \vert M,\dl) d\dl'' = 
\sum_{i=1}^n \dfrac{M'_i}{M} \ d\Nc_i\ \phi( M'_1,\dl'' \vert M'_i,\dl') d\dl'',
\nonumber
\end{equation}
from which we derive the following integral equation: 
\begin{eqnarray}
\nonumber
& & \phi( M',\dl'' \vert M,\dl) d\dl''=   \\  
& & = \mbox{} \int_{M'}^M dM''\ \dfrac{M''}{M}\ 
\nonumber
\Nc( M'',\dl' \vert M,\dl)\ 
\phi( M',\dl'' \vert M'',\dl') d\dl''
\end{eqnarray}
By integration on $\dl''$ between $\dl'$ and $\infty$, the previous equation writes in terms of the 
cumulative field conditional probability:
\begin{eqnarray}
\label{fundeqa}
& & F( M',\dl' \vert M,\dl) =   \\
& &= \mbox{} \int_{M'}^M dM'' \dfrac{M''}{M}\ 
F( M',\dl' \vert M'',\dl')\
\Nc( M'',\dl' \vert M,\dl) 
\nonumber 
\end{eqnarray}

An alternative formulation of Eq. (\ref{fundeqa}) is obtained introducing the {\it conditional probability} 
$p(M', \dl' \vert M, \dl)dM'$
of finding an isolated region of mass between $M'$ and $M'+dM'$ and overdensity $\dl'$ inside a region $\bf S$. 
Given a region $\bf S$ with overdensity $\dl$ and mass $M$, our interest is now to 
identify inside it an isolated region with overdensity $\dl'$.
The quantity
\begin{equation*}
\Delta M'' = M'' \Nc(M'',\dl' \vert M,\dl)~dM''
\end{equation*} 
gives the amount of mass of $\bf S$ assembled by isolated regions with overdensity $\dl'$ and mass between
$M''$ and $M'' + dM''$.
Therefore, the probability that an arbitrary mass element of $\bf S$ is part of an isolated region with
overdensity $\dl'$ and mass between $M''$ and $M'' + dM''$ is
\begin{equation*}
p(M'', \dl' \vert M, \dl)~dM'' = \dfrac{\Delta M''}{M},
\end{equation*}
that is
\begin{equation}
\label{prob}
p(M'', \dl' \vert M, \dl) = 
\dfrac{M''}{M}~\Nc(M'',\dl' \vert M,\dl) .
\end{equation}

The previous statement is equivalent to say that $p(M'', \dl' \vert M, \dl)~dM''$ is the probability to 
find an isolated region with overdensity $\dl'$ and mass between $M''$ and $M'' + dM''$  inside $\bf S$.
Introducing Eq. (\ref{prob}) into Eq. (\ref{fundeqa}), we obtain
\begin{eqnarray}
\label{fundeqb}
& & F( M',\dl' \vert M,\dl) =   \\
& &= \mbox{} \int_{M'}^M dM''\ 
F( M',\dl' \vert M'',\dl')\
p( M'',\dl' \vert M,\dl) 
\nonumber 
\end{eqnarray}
which is basically a Volterra integral equation.

The fundamental aspect of Eq. (\ref{fundeqa}) is that for any given cumulative field conditional 
probability $F( M' , \dl' \vert M , \dl)$, it gives the conditional MF, 
$\Nc( M' , \dl' \vert M , \dl)$.
Its main difference with the Jedamzik (1995) result is that Eq. (\ref{fundeqa}) is applied to any
region $\bf S$ with overdensity $\dl$ and mass $M$ and not only to the overall universe.  
For this reason the cumulative field conditional probability $F( M' , \dl' \vert M , \dl)$ 
appears now on both sides of the equation.
Such property highlights the intrinsic meaning of the CF.
It is interesting to remark that the previous {\it isolated overdense regions inventory} theorem solves 
automatically the cloud-in-cloud problem.
The Eq. (\ref{fundeqa}) represents the generalization of the PS formalism. 
Such formalism may be recovered for the special random fields  wherein  
\begin{equation}
F( M' , \dl \vert M , \dl) = k 
\label{anscon}
\end{equation} 
with $k$ constant.   
In such case Eq. (\ref{fundeqa}) writes
\begin{equation}
\label{ans}
p(M',\dl' \vert M,\dl)= 
-\dfrac{1}{k} \dfrac{dF(M',\dl' \vert M,\dl)}{dM'}
\end{equation} 
The PS Ansatz in terms of the conditional probability and including already the fudge factor of 2, 
is obtained in the particular case of $k=1/2$ (corresponding namely to Gaussian fields, see below). 

A random field for which Eq. (\ref{anscon}) is satisfied is very simple to handle in terms of Eq. (\ref{ans}). 
Note that not only Gaussian fields obey Eq. (\ref{anscon}) but also a variety of non-Gaussian fields. 
Therefore, in order to find an analytical formalism in agreement with simulations and to avoid the 
complex numerical integration of Eq. (\ref{fundeqa}), it is natural for us to look at first among 
random fields which satisfy Eq. (\ref{anscon}).
This is the strategy followed in Section 4.

\subsection{Merger trees build up}

A merger tree realization needs to identify progenitors of a given descendant. This is done 
by means of a Monte Carlo algorithm that uses the conditional probability and MF.
The choice of a Monte Carlo algorithm is not unique and may hide subtle physical 
and technical questions (for a review see Zhang et. al. 2008). The simplest criterion 
establishes that the Monte Carlo algorithm has to reproduce the conditional probability 
and MF predicted by the theory.
However, this criterion is not sufficient, because the overall meaning of the involved probability,
i.e. the stocahstic nature of the problem, has to be taken in to account. 
We will introduce a Monte Carlo algorithm optimized to build up a merger tree.
Afterwards we will check our algorithm with tests that involve the conditional probability and
MF, as well as the MAH and MR in order to verify also the agreement with the stochastic nature 
of the problem.

Only in this section, to ease the reading of the formulae, we will indicate the variables of the 
extracted regions by a lower case letter.

Integrating $p(m', \dl' \vert M, \dl)dm'$ between $0$ and $m$ and identifying the obtained cumulative 
probability with a random number $\alpha$ uniformly distributed in the interval [$0$--$1$], we write 
\begin{equation}
\label{Mont}
\alpha = \int_0^m~p(m',\dl' \vert M,\dl)~dm',
\end{equation} 
which represents the basis for a Monte Carlo approach to build up merger trees.

Given an isolated region ${\bf I_0}(\dl_0)$ with overdensity $\dl_0$ and mass $M_0$, the extraction of the random 
number $\alpha_1$ by Eq. (\ref{Mont}) allows to identify inside it a first isolated region ${\bf i_1}(\dl')$ of 
mass $m_1$ with a given overdensity $\dl'$. 
Taking into account that the {\it isolated overdense regions inventory} theorem has been proved for general regions
(see footnote$^2$), after the identification of the isolated region ${\bf i_1}(\dl')$, the extraction may 
continue on the complementary region $\bf S_1(\dl_1)$ with mass $M_1=M_0-m_1$ and volume $V_1=V_0-v_1$ 
obtained from ${\bf I_0}(\dl_0)$ removing ${\bf i_1}(\dl')$ from its inside. 
In terms of the density, from the volume relation we write
\begin{equation*}
\dfrac{M_0}{\rho_0} = \dfrac{m_1}{\rho'}+\dfrac{M_1}{\rho_1},
\end{equation*}
Being the region density $\rho =\rbh (\dl+1)$, where $\rbh$ is the background density and $\dl$ the
overdensity, we have
\begin{equation*}
\dfrac{M_0}{\dl_0+1} = \dfrac{m_1}{\dl'+1} + \dfrac{M_1}{\dl_1+1},
\end{equation*}
and finally 
\begin{equation*}
\dl_1 = \dfrac{M_0-m_1}{\dfrac{M_0}{\dl_0+1} - \dfrac{m_1}{\dl'+1}} - 1.
\end{equation*}

At this point the extraction of the next isolated region can be carried out on the region $\bf S_1(\dl_1)$ 
with overdensity $\dl_1$ and mass $M_1$.
To perform such extraction, for simplicity, we will use Eq. (\ref{Mont}) with a new random number $\alpha_2$,
applied on the region $\bf S_1(\dl_1)$.
By proceeding in this way we implicitly assume that the same statistics of ${\bf I_0}(\dl_0)$ holds for 
$\bf S_1(\dl_1)$, excluding any sort of correlation.
For a discussion of possible correlations, see Sheth \& Lemson (1999).
According to what has been said above, afterwards we will proceed to test the robustness of such hypothesis.

A recurrent algorithm may be particularly useful now. 
From the $n^{th}$ complementary region $\bf S_n$ of mass $M_n$ with overdensity $\dl_n$ calculated by 
the previous method, using Eq. (\ref{Mont}) with a random number $\alpha_{n+1}$ we can identify inside 
it an isolated region ${\bf i_{n+1}}(\dl')$ with mass $m_{n+1}$ and overdensity $\dl'$. 
At this point the next complementary region $\bf S_{n+1}$ with mass
\begin{equation}
\label{mcon}
M_{n+1}=M_n-m_{n+1}
\end{equation} 
will have the overdensity
\begin{equation}
\label{overev}
\dl_{n+1} = \dfrac{M_n-m_{n+1}}{\dfrac{M_n}{\dl_n+1} - \dfrac{m_{n+1}}{\dl'+1}} - 1.
\end{equation} 
Equations (\ref{mcon}) and (\ref{overev}) guarantee a mass conservation and make the procedure 
particularly handy because they do not make explicit reference to the previous i-steps.

Equations (\ref{Mont}--\ref{overev}) are particularly useful for constructing the halo merger 
trees. 
Suppose a simple case when the isolated virialized regions are defined by a critical overdensity $\dl_c$ 
function of $z$ only. Given a {\it descendant} at $z$ with overdensity $\dl_c(z)$, the {\it progenitors} 
at $z+ \Delta z$ have overdensity $\dl_c(z+ \Delta z)$. The finding of such progenitors (branches) is 
performed by applying the Monte Carlo extractions of Eq. (\ref{Mont}) to the (complementary) region 
described by Eqs. (\ref{mcon}) and (\ref{overev}). 
This procedure identifies a sequence of progenitors at $z+ \Delta z$ that can be sorted in order of decreasing mass. 
The largest (primary) progenitor identifies the main progenitor by which the {\it mass aggregation history,} MAH, 
is obtained. The overall population of less massive progenitors fulfills the {\it merger tree} and allows to find 
the {\it merger rate}, MR, along the MAH.

Note that the condition $\rho = 0$ implies $\dl = -1$. 
Therefore, for the complementary regions, the progenitor extractions can be performed meanwhile $\dl_n \geq -1$.
This means that the progenitors are extracted until all matter in the volume is exhausted, which implies that
the entire mass contained in the descendant should come from virialized structures.
However, the physics of halo mass assembly is more complex. 
A fraction of the mass can be present as diffuse matter. 
The strict limit $\dl \geq -1$ implies lack of diffuse matter.
In this paper we explore solutions with $\dl \geq -1$, however we will have in mind the physical consequences of 
such a hypothesis.  

The above described method has been extensively used in the case of Gaussian CDM density 
fields for generating the initial conditions (the MAHs) for the virialization process of dark haloes (Avila-Reese, 
Firmani \& Hernandez 1998), and the subsequent formation and evolution of disc galaxies inside them (Firmani \& 
Avila-Reese 2000,2009; Avila-Reese \& Firmani 2000).
The obtained halo properties agree with those found in cosmological numerical simulations (Avila-Reese et 
al. 1999).

It is interesting to remark that the mass distribution of the first progenitor obtained with
the Monte Carlo method can be compared with the conditional probability function, while
the mass distribution of all the progenitors can be compared with the conditional MF. 
This offers a first test of our Monte Carlo algorithm by comparing its one-step results with the 
straight analytic formulae from the statistics (in \S\S 3.2 we will show this point in some detail).

It is important to highlight that the mean halo MAH has to be a convergence solution of the Monte Carlo
algorithm for the integration redshift step $\Delta z \rightarrow 0$, otherwise it is physically meaningless.
Because of the stochastic nature of the problem, this property is dependent on the behavior of the given 
cumulative field conditional probability. We will focus on {\it well behaved cumulative field
conditional probabilities, such that guaranty $\Delta z \rightarrow 0$ convergence in the MAH buildup}.

\section{Gaussian random fields}

In the case of a Gaussian random field, the cumulative field conditional probability to find a region 
with mass $M'$ and overdensity $\geq \dl'$ contained inside the region of mass $M$ with overdensity $\dl$ is 
given (see Bower 1991) by:
\begin{equation}
\label{gaussdist}
F(M',\dl' \vert M,\dl)=\dfrac{1}{\sqrt{\pi}}
\int_{\small{\gamma}}^{\infty}~e^{- \xi^2} d\xi,
\end{equation}
where
\begin{equation}
\label{gamma}
\gamma = \dfrac{\left( \dl' - \dl \right) }{\sqrt{2 \left( \sg^2_{M'} - \sg^2_M \right) }},
\end{equation} 
and $\sg_M$ is the mass variance calculated from the density fluctuation power spectrum of the random field. 
Making the limit $\dl' \rightarrow \dl$,  Eq. (\ref{gaussdist}) gives
\begin{equation}
\label{gaussianval}
F(M',\dl \vert M,\dl) = \dfrac{1}{2}~~~~~~~~(M'<M),
\end{equation}
which is Eq. (\ref{anscon}) with $k=1/2$. Then Eq. (\ref{ans}) reduces to
\begin{equation}
p(M',\dl' \vert M,\dl)= 
 -2\dfrac{dF(M',\dl' \vert M,\dl)}{dM'}.
\label{psans}
\end{equation}
This is just the PS Ansatz in terms of the conditional probability with the fudge factor of 2 included. 
Introducing Eq. (\ref{gaussdist}) into Eq. (\ref{psans}) we obtain 
\begin{equation}
\label{psgen}
p(M', \dl' \vert M, \dl) =
\dfrac{2}{\sqrt{\pi}}~e^{- \gamma^2}~\dfrac{d\gamma}{dM'}.
\end{equation}
It is easy to verify that such conditional probability is normalized, i.e., its integral in $M'$ from $0$
to $M$ ($\gamma$ from $0$ to $\infty$) is 1.

\subsection{Unconditional mass function for haloes}

Eq. (\ref{psgen}) is applied to a region of mass $M$ and overdensity $\dl$. 
We can extend such region to the entire universe making $M \rightarrow \infty$, $\sg_{M} \rightarrow 0$, 
and $\dl \rightarrow 0$.
Hereafter, in such limit case of the entire universe, the superscript $'$ will be omitted. 
A recipe for the collapse criterion may be obtained from the top-hat spherical gravitational collapse, 
which establishes that an isolated region at redshift $z$ is virialized when its overdensity is equal to a 
critical value $\dl_c$ (Navarro Frenk \& White 1997). 
For a flat cosmology with cosmological constant ($\Omega_M + \Omega_\Lambda =1$): 
\begin{equation}
\label{dlc}
\dl_c=\dfrac{1.686 \ \Omega_M^{0.0055}}{D^+(z)}
\end{equation} 
where $D^+(z)$ is the linear growth factor normalized to 1 at $z=0$.

By defining the number density per mass unit
\begin{equation}
\label{cmf}
\dfrac{dN}{dM} \equiv \dfrac{\rbh}{M}\ p(M,\dl_c \vert \infty, 0),
\end{equation} 
definition that is not limited to Gaussian fields, and introducing
\begin{equation}
\label{nu}
\nu = \dfrac{\dl_c}{\sg_M},
\end{equation} 
we obtain the halo MF
\begin{equation}
\label{ps}
\dfrac{M}{\rbh}~\dfrac{dN}{dM}=\sqrt{\dfrac{2}{\pi}}~e^{- \frac{\nu^2}{2}}~\dfrac{d\nu}{dM} 
\end{equation}
which coincides with the PS MF.

\begin{figure}
\includegraphics[scale=0.55, angle=-90]{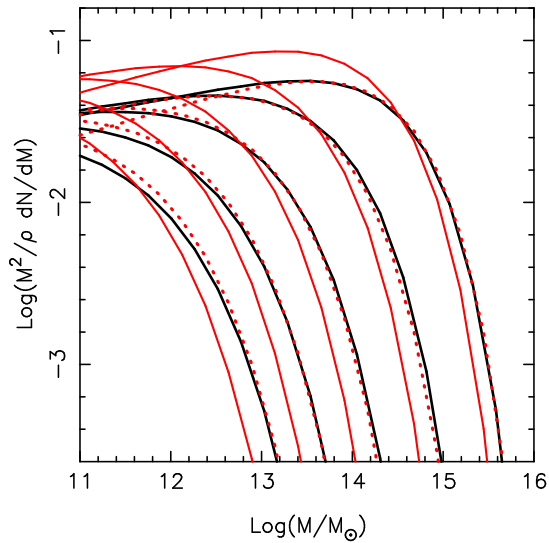} 
\caption{Halo MFs plotted as $(M^2/\rbh)dN/dM$ from Tinker08 (solid black lines)
compared to the PS MFs obtained with the CF (solid red lines) for $z=0,1,2,3,4$.
The dotted red lines show the PS MFs with rescaled critical overdensity and lowered 
amplitude (see text).}
\label{PSmafun}
\end{figure}

In Fig. \ref{PSmafun}, the solid red curves show the Gaussian-case (PS) halo MF 
obtained from Eq. (\ref{ps}) at $z=0,1,2,3,4$, while black solid curves are the accurate fitting 
formulae to cosmological N-body simulations provided by Tinker et al. (2008; hereafter Tinker08); 
the parameters for the virial mass at the overdensity $\Delta(z)$ (Bryan \& Norman 1998) corresponding 
to our cosmology were used. 
The comparison reveals the well known excess (deficit) of intermediate (high) mass haloes predicted by 
the PS formalism. 
The Tinker08 fitting curves are valid for halo masses above $\sim 10^{11} h^{-1} M_{\odot}$, which 
corresponds to the minimal mass at which haloes are reasonable resolved in the simulations studied 
by these authors; this limitation makes uncertain the normalization of the overall MF. 
Just in order to explore possibilities to approximate the PS MF to the simulations MF, $\dl_c$ has 
been rescaled by a factor $0.86$ to fit the high mass cut-off of the N-body (Tinker08) curves 
(for similar earlier attempts see e.g., Carlberg \& Couchman 1989; Klypin \& Rhee 1994), 
and the normalization has been lowered by a factor $0.61$ to fit the maxima (red dotted curves).  
Note that the PS MF is normalized independently of $\dl_c$, and that the dotted curves imply now a not 
normalized MF. 
It is evident that in spite of these mass-independent transformations of the PS MF, the comparison with 
simulation results continue failing. 
The exercise presented above is congruent with the one carried out in Sheth \& Tormen (1999), where they 
showed that besides these operations, the PS MF should be multiplied by a $\nu$ (mass)-dependent factor 
in order to fit the numerical simulations analyzed by them.

\subsection{Mass aggregation histories and merger rates}

For a Gaussian field and the critical overdensity given by Eq. (\ref{dlc}), 
if $M' \rightarrow 0$, then $\sg_{M'} \rightarrow \infty$, $\gamma \rightarrow 0$ and 
$F(M',\dl' \vert M,\dl) \rightarrow 1/2$.
Using Eqs. (\ref{psans}) and (\ref{gaussdist}), Eq. (\ref{Mont}) writes as
\begin{equation}
\label{pro}
\alpha = \dfrac{2}{\sqrt{\pi}} \int_{0}^{\gamma} e^{- \xi^2} d \xi .
\end{equation}
This operation is equivalent to make a random choice of a number $\gamma$ with a normal 
deviate, zero mean and variance $0.5$. 
Hence, using Eq. (\ref{gamma}) 
and making reference to \S\S 2.2, we obtain:
\begin{equation}
\label{meq}
\sg_{M'_i}^2 = \dfrac{\left( \dl'_i-\dl_i \right)^2}{2 {\gamma}^2} + \sg_{M_i}^2,
\end{equation}
where the symbols are the same as in \S\S 2.2, and $\dl_0=\dl_c(z)$ is the (rescaled critical) overdensity
of the descendant at redshift $z$, while $\dl'_i=\dl_c(z+\Delta z)$ is the (rescaled critical) overdensity
of a progenitor at redshift $z+\Delta z$. 
The $i$ sequence defines the complementary regions of mass $M_i$ with overdensity $\dl_i$, as well as the
progenitors of mass $M'_i$ with overdensity $\dl'_i$.
Equations (\ref{mcon}), (\ref{overev}) and (\ref{meq}), togheter, define the Monte Carlo algorithm to build 
the merger tree of a virialized halo with a given mass at a given redshift.  

\begin{figure} 
\includegraphics[scale=0.55, angle=-90]{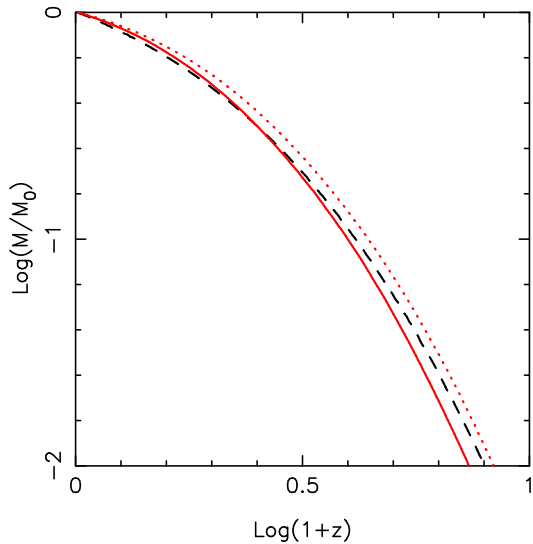}
\caption{Average MAH (solid red line) for a $M_0 = 10^{13} M_{\odot}$ 
present-day halo for the PS case, and the same for the amplitude-reduced case with rescaled critical 
overdensity (dotted red line, see text).
The dashed black line shows the corresponding mean MAH obtained from the
Millennium Simulations (Fakhouri et al. 2010).}
\label{PSmafr}
\end{figure}

\begin{figure}
\includegraphics[scale=0.55, angle=-90]{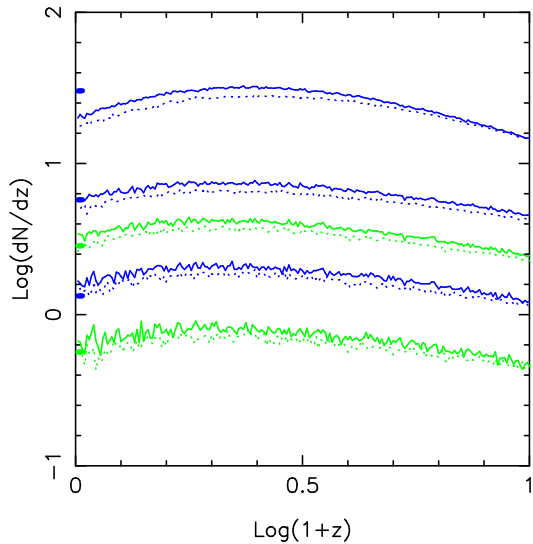} 
\caption{Average MRs for the PS case (from bottom to up, green lines for $\xi>0.3$ 
and $>0.03$, blue lines for $\xi>0.1$, $0.01$ and $0.001$) along the average MAHs of  
Fig. \ref{PSmafr} (solid lines for the main case and dotted lines for the amplitude-reduced case 
with rescaled critical overdensity, see text). 
Taking into account the mass variation, this result compares well with the Millennium 
Simulations at $z\approx 0$ (the thick in the vertical axis indicate the corresponding $z=0$ MRs
reported in Fakhouri et al. 2010). 
The evolutionary behavior also agrees with these simulations.
}
\label{PSmefr}
\end{figure}

In Fig. \ref{PSmafr}, we plot the average of $2\ 10^4$ different MAHs corresponding to a present-day halo
of mass $M_h=10^{13} M_{\odot}$ (red solid line; in this case, the critical overdensity is not rescaled). 
As in Fig. \ref{PSmafun}, the red dotted line corresponds to the rescaling in the 
critical overdenstiy and the amplitude-reduction correction of the MF mentioned above. 
The dashed line represents the fit to the corresponding average MAH measured in the Millennium Simulations 
(Fakhouri, Ma \& Boylan-Kolchin 2010).

For present-day haloes of $M_h=10^{13} M_{\odot}$, Fig. \ref{PSmefr} shows the mean MR histories 
{\it per unit of redshift} 
with merger ratios greater than $\xi\equiv M_s/M_p$ 
($M_p$ and $M_s$ are the masses of the primary and secondary progenitors, respectively). 
From bottom to top, $\xi > 0.3$ and $0.03$ (green lines) and $\xi > 0.1$, $0.01$, and $0.001$ (blue lines). 
The coding of solid and dotted lines is the same as in Fig. \ref{PSmafr}. 
The average halo MRs as a function of the $\xi$ threshold at $z=0$ compare well with the results reported in 
Fig. 3a by Fakhouri et al. (2010). 
The small change of these mean MRs with $z$ is also in general agreement with the numerical simulation results
reported in Fakhouri \& Ma (2008) and Fakhouri et al. (2010).

\begin{figure}
\hspace{6.0 mm}
\includegraphics[scale=0.325]{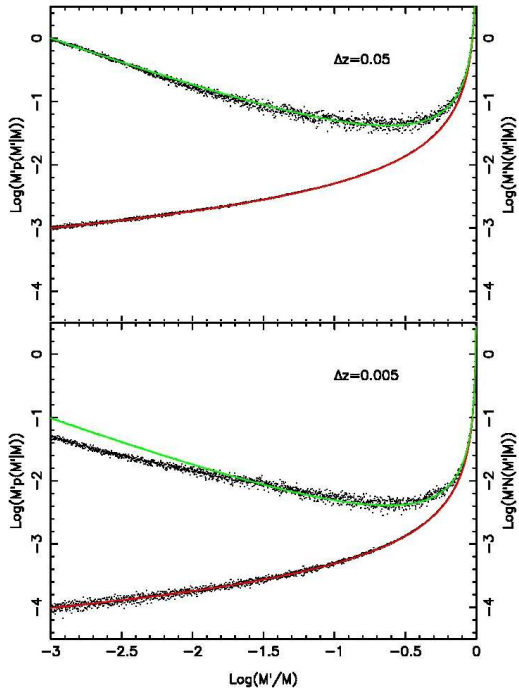} 
\caption{The graphic shows the conditional probability $p( M',\dl' \vert M,\dl) $ (red, left abscissa) 
and the conditional mass function $\Nc( M',\dl' \vert M,\dl) $ (green, right abscissa) 
per unitary progenitor mass natural logarithm as a function of the progenitor mass. 
The dots show the average distribution obtained by the Monte Carlo algorithm for the first progenitor (bottom) and 
for the entire progenitor collection (top) using Eqs. (\ref{mcon}) and (\ref{overev}). 
The descendant mass is $10^{13} \msun$, the redshift $z=0$ while the redshift step $\Delta z = 0.05$ (upper panel) 
and $\Delta z = 0.005$ (lower panel).}
\label{test1}
\end{figure}

\begin{figure}
\hspace{6.0 mm}
\includegraphics[scale=0.325]{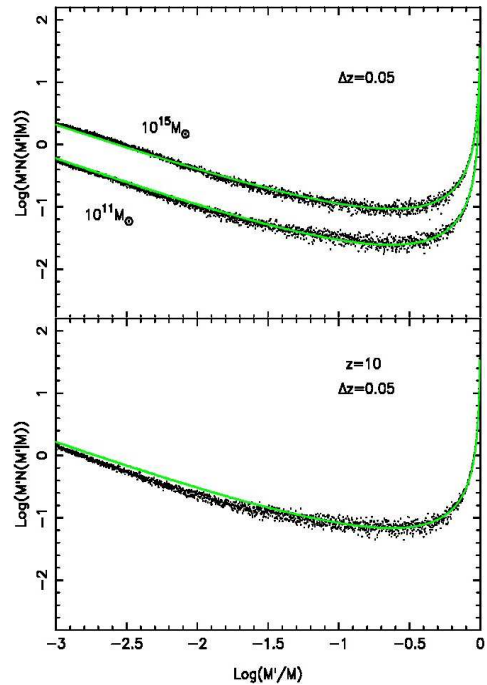} 
\caption{The graphic shows the conditional mass function $\Nc( M',\dl' \vert M,\dl) $ (green) per unitary progenitor 
mass natural logarithm as a function of the progenitor mass. 
The dots show the average distribution obtained by the Monte Carlo algorithm for the entire progenitor collection 
using Eqs. (\ref{mcon}) and (\ref{overev}). 
In the upper panel the descendant masses are $10^{11} \msun$ and $10^{15} \msun$, the redshift $z=0$ and the redshift 
step $\Delta z = 0.05$, while in the lower panel the descendant mass is $10^{13} \msun$, the redshift $z=10$ and the
redshift step $\Delta z = 0.05$.}
\label{test2}
\end{figure}

The main conclusion from the exercise presented here is that for a Gaussian density field (PS),
the predicted average halo MAH and MRs agree reasonably well with the results from numerical 
simulations
\footnote{For the amplitude-reduced case with rescaled critical overdensity, the MAH and MRs 
do not change significantly (Figs. \ref{PSmafr} and \ref{PSmefr}). 
In what follows,  unless otherwise stated, we will use the rescaled critical overdensity case.}.
However, the predicted halo MF, as it is well known, has an excess at intermediate masses and a deficit 
at high masses as compared to simulations at the mass range they are able to resolve (e.g., Tinker08).

Particularly interesting are the one-step results shown in Fig. \ref{test1} for a descendant mass
of $10^{13} \msun$ at $z=0$ and redshift intervals of $\Delta z = 0.05$ (upper panel) and $\Delta z = 0.005$
(lower panel), respectively.
The red curve (left abscissa) shows the analytic conditional probability $p( M',\dl' \vert M,\dl) $ per 
unitary progenitor mass natural logarithm, while the green curve (right abscissa) shows the conditional MF 
$\Nc( M',\dl' \vert M,\dl) $ per unitary progenitor mass natural logarithm as a function of the progenitor mass.
Here $\dl$ correspond to $z=0$ while $\dl'$ to $z=\Delta z$. The dots show an average distribution 
obtained with our Monte Carlo algorithm for the first progenitor (bottom region) and for the entire 
progenitor collection (top region). The total number of extractions, not reported here because unnecessary, 
is regulated in each case to reduce the scatter at reasonable levels. The marginal differences appearing in 
the graphics can arise from numerical effects, as well as from some departure from the hypothesis  
made in \S\S 2.2 about the complementary regions. 
Fig. \ref{test2} shows a similar test carried out for the masses $10^{11} \msun$ and $10^{15} \msun$ at $z=0$
(upper panel) and for $10^{13} \msun$ but at $z=10$ (lower panel), being in both cases $\Delta z = 0.05$.
We conclude that the agreement between the analytic conditional probability and MF and the Monte Carlo results 
as well as the comparison of the MAHs and MRs with simulations (see above) are fully satisfactory, and it indicates that 
our merger tree algorithm works properly. For our main calculations we will assume a redshift step close to 
$0.05$, while for step invariance tests we will assume $0.005$.

\subsection{The effect of mass dependence in $\delta_c$}

\begin{figure}
\includegraphics[scale=0.55, angle=-90]{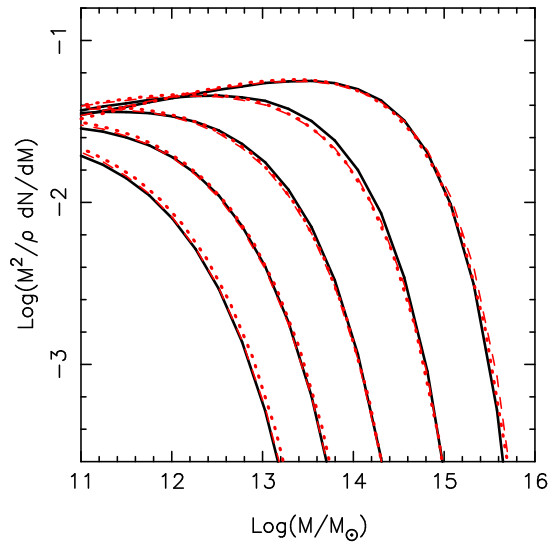} 
\caption{Halo MFs plotted as $(M^2/\rbh)dN/dM$ from Tinker08 (solid black lines) compared with the MFs obtained 
with the CF for $\dl_c$ dependent on mass (dashed red curves) for $z=0,1,2,3,4$.
The dotted red lines show the Sheth-Tormen (1999) MFs.}
\label{STMF}
\end{figure}

So far, we have used the spherical collapse critical overdensity (Eq. \ref{dlc}), which is function of $z$ but not
of mass; this corresponds to a {\it fixed barrier} in the ES formalism. 
The adoption of a mass-dependent critical overdensity, i.e. a {\it moving barrier,} compatible 
with an ellipsoidal collapse, improves the agreement between the analytic and the numerical results regarding 
the halo (unconditional) MF (Sheth et al. 2001; Sheth \& Tormen 2002). 
We extend now the test to the halo MAH and MR.
We assume the same Gaussian random field as in Section 3 described by Eqs. (\ref{gaussdist})--(\ref{psgen}) 
and Eqs. (\ref{cmf})--(\ref{ps}), and adopt the moving barrier
\begin{equation}
\label{dlec}
\dl_{ec}=a\ \dl_c\ \left( 1 + f(\chi)\right),
\end{equation}
where $\chi = \sigma_M/\dl_c$, $f$ is a function that in Sheth et al. (2001) is $f=b\chi^c$, and 
$a$, $b$, and $c$ are free parameters. 
The CF can be used now to explore the effects of a mass-dependent critical overdensity on the MF, MAH and MR,
under the Anstaz that the cumulative field conditional probability is the same as given in eq. (\ref{gaussdist}).
Using Eq. (\ref{ps}) with the parameter values given in Sheth et al. (2001), $a=0.87$, $b=0.47$ and $c=1.23$, 
we recover almost the same halo MF proposed by Sheth \& Tormen (1999) as a fit to numerical simulations, 
and justified in Sheth et al. (2001) as an indication of elliptical collapse (Fig. \ref{STMF};
note that both the obtained MF and the Sheth \& Tormen one are close to the MF given in Tinker08). 
The excellent agreement shows that our Ansatz is quite reasonable.
Within the framework of the ES approach, recent work on random walks with correlated steps suggests that 
substituting $\dl_{ec}$ into Eq. (\ref{gamma}) could be indeed a good approximation (c.f.  Paranjape, 
Lam \& Sheth 2012; Musso \& Sheth 2012). 

To find the MAH and MR, we use Eqs. (\ref{gaussdist})--(\ref{psans}) with $\dl_{ec}$ given by Eq. (\ref{dlec}).
For $M' \rightarrow 0~$ ($\sigma_{M'} \rightarrow \infty$), $\gamma \rightarrow 0$ only if $c < 1$, while 
$\gamma \rightarrow ab/\sqrt{2}$ when $c = 1$ and $\gamma \rightarrow \infty$ for $c >1$. 
In these three cases, $\varphi\equiv F(0,\dl'_{ec} \vert M,\dl_{ec})$ is $1/2$, a value between $0$ and $1/2$, and $0$,
respectively.
Using Eq. (\ref{psans}), Eq. (\ref{Mont}) now writes
\begin{equation}
\label{proe}
\alpha - 2 \varphi + 1 = \dfrac{2}{\sqrt{\pi}} \int_{0}^{\gamma} e^{- \xi^2} d \xi  
\end{equation}
This equation does not have a solution when $c > 1$, while it does have a solution when $c < 1$ for each value of 
$0 < \alpha < 1$, and for $c = 1$ only if $2 \varphi - 1 < \alpha < 2 \varphi$.
Within the framework of our approach, we conclude that 
a mass-dependent critical overdensity, as the one inferred from the moving barrier analysis suggested in Sheth et. al (2001),
does not offer a consistent description of the halo MAH and MR, though, 
the limiting "square-root" moving barrier ($c = 1$) suggested by Moreno et al. (2008) in principle 
could offer a solution. 

The results obtained by the Monte Carlo method using the {\it square-root} moving barrier show that both the 
average halo MAH and MR depart significantly from those found in the simulations. 
The MAH is now too extended towards the past, revealing a low contribution of major mergers in the halo assembly. 
The major MRs ($\xi>0.1$) are indeed rare with respect to the simulation results.
We have explored the obtained effects varying the $a,b,c$ parameters without any interesting result.

Nonetheless, the most serious difficulty in our formalism when using a mass-dependent critical overdensity
is that the convergence of the Monte Carlo algorithm, when redshift step $\Delta z \rightarrow 0$, does not 
accomplish anymore.
It is not surprising that some rough agreement with the results from simulations is obtained only when 
$\Delta z \gtrsim 0.5$, owing to the "correct" behavior of $\gamma$ as a function of the progenitor mass 
in this case (see below).
Indeed, by using the large  $\Delta z \gtrsim 0.5$ time step, Moreno et al. (2008) have showed that
the moving barrier predictions agree with the simulation conditional probability. 

\begin{figure}
\includegraphics[scale=0.325, angle=-90]{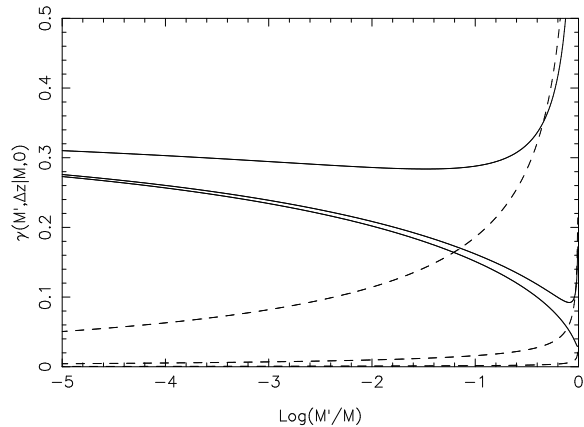} 
\caption{The plot shows $\gamma$ as function of the mass in the case of a fixed barrier
(dashed lines) and a moving barrier (solid lines). The descendant mass is $10^{13}M_\odot$,
$\Delta z = 0.005, 0.05, 0.5$ from bottom to top}
\label{Gamplot}
\end{figure}

Simple considerations help us to understand the failure of the mass-dependent critical overdensity
in our scheme. Figure \ref{Gamplot} allows to understand the reduction of the major merger rate.
For a fixed barrier ($\dl_c$ independent of mass; dashed lines), $\gamma$ increases monotonically 
with mass and Eq. (\ref{psgen}) leads to a conditional probability that makes possible the existence of low-mass progenitors. 
For the "square-root" moving barrier (Moreno et al. 2008; solid lines), $\gamma$ shows a decreasing 
shape on a very extended low-mass range of progenitors; here the conditional probability Eq. (\ref{psgen}) 
is physically meaningless and no low mass progenitors are possible.
Only in a small range immediately below the descendant mass, $\gamma$ increases with mass, allowing
at this range for a physical conditional probability, such that the progenitors are more massive than in
the fixed barrier case. 
When $\Delta z > 0.5$ the "correct" monotonic increasing of $\gamma$ with the progenitor mass is recovered.
Furthermore, a necessary condition to obtain a redshift-step convergence is that $\gamma \rightarrow 0$
when $\Delta z \rightarrow 0$ with $M' < M$. This tendency is not fulfilled for the moving barrier case.

We conclude that, {\it when using a mass-dependent critical overdensity (moving barrier) in our approach,
the halo MF is correctly described but the predicted halo MAH and MR fail in reproducing the simulation results.}
An exhaustive analysis of this shortcoming and the reason behind it is deserved for a future work.

\section{A semi-empirical strategy} 

In view of the difficulty of proposing a random fluctuation field described by an analytic cumulative 
field conditional probability function and being able to generate at the same time the halo MF, MAH, and MR
obtained in the numerical simulations, we turn on to an inductive (semi-empirical) approach: 
the probability function will be inferred numerically from the Tinker08 halo MF, which was obtained from 
simulations. We split the problem into two parts. Firstly, in \S 4.1 we extend the unconditional 
probability corresponding to the Tinker08 MF (defined only above $\sim 10^{11}$ \msun) down to 
small masses by an algebraic extrapolation, taking care that such an extrapolation does not imply 
a sharp change in the functionality; furthermore, we introduce a mass rescaling in order to take into 
account any halo mass defect due to some diffuse matter (see \S\S 2.2).
Secondly, in \S 4.2, based on the properties of Eq. (\ref{fundeqa}), we extend the definition of the unconditional 
probability to the conditional probability, and using it within the context of the CF, we calculate the halo MAH and
MRs. 

\subsection{Halo mass rescaling in the numerical simulations}

Tinker08 provided an analytical fit to the halo MFs measured in numerical simulations at different redshifts
and above a minimum (resolution) mass. We can calculate from these MFs the cumulative unconditional probability 
functions expressed in terms of the commonly used scaled variable  $\nu$ or, what is the same, 
$\gamma_u = \nu / \sqrt{2} = \dl_c(z)/\sqrt{2}\sigma_M$, where
$\gamma_u$ is the same as in Eq. (\ref{gamma}) but applied to the entire universe, i.e. for
$M\rightarrow \infty$ ($\sigma_M\rightarrow 0$) and $\delta\rightarrow 0$. 
Due to the lower limit in mass in the Tinker08 MFs, the cumulative unconditional probability should be 
constructed by integrating the MF from each mass $\tilde{M}$ at a given redshift, corresponding to the 
argument $\gamma_u$, to infinity. 
We define the following function:
\begin{eqnarray}
\label{Tpro}
& & P(\gamma_u;z) \equiv 1 - \int_{\tilde{M}(\gamma_u;z)}^\infty \dfrac{M}{\rbh}~\dfrac{dN}{dM}~dM =   \\
& &=  \int_0^{\tilde{M}(\gamma_u;z)}~p(M ,\dl_c(z) \vert \infty, 0)~dM  
\nonumber  
\end{eqnarray}
where $dN/dM$, given by Eq. (\ref{cmf}), is the halo MF,
and the mass $\tilde{M}(\gamma_u;z)$ is an intrinsic function of $\gamma_u$ for a given $z$.
As long as the integral in Eq. (\ref{Tpro}) is correctly normalized, $P(\gamma_u;z)$ represents the cumulative 
probability to find an isolated region with overdensity $\dl_c(z)$ (collapsed region) and mass $\leqslant \tilde{M}$
(or $\gamma_u$ less than the value corresponding to $\tilde{M}$ for the given $z$). 
When we apply Eq. (\ref{psans}) to the Gaussian (PS) case, we obtain 
$P_{PS}(\gamma_u;z) = 1-2 F(\tilde{M},\dl_c \vert \infty, 0)$.
Since in such case $F$ is function of $\gamma_u$ only,  
the dependence of $P_{PS}(\gamma_u;z)$ on $z$ vanishes ($P_{PS}$ depends on $z$ but through 
$\gamma_u$); this is the well known fact that for a Gaussian field, the halo multiplicity function is 
universal when expressed in terms of the scaled variable $\nu$ (or  $\gamma_u$). 

\begin{figure} 
\hspace{-3.5 mm}
\includegraphics[scale=0.37]{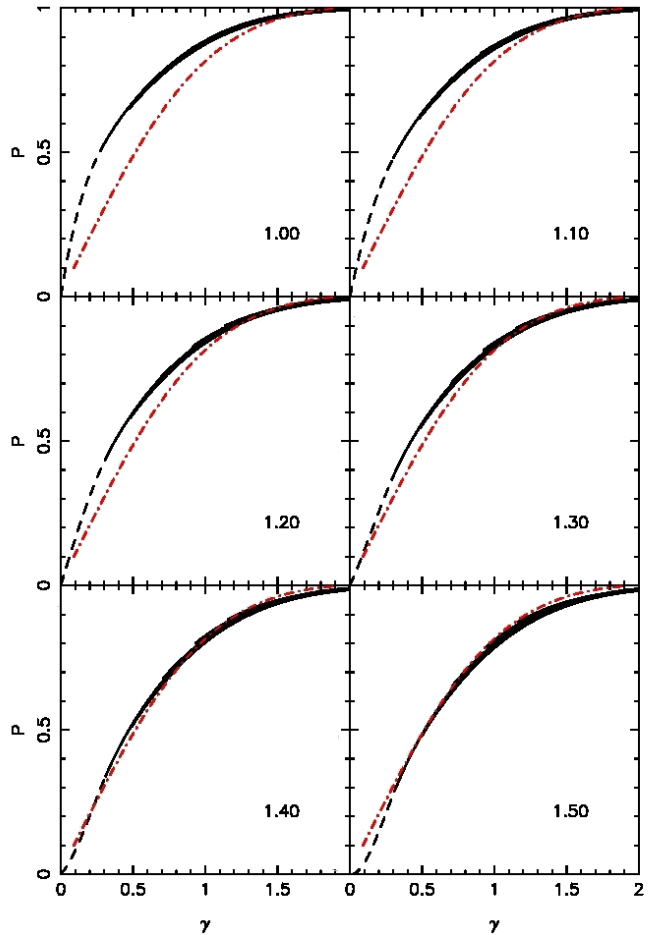}
\caption{Back integrated cumulative probability $P$ vs. $\gamma_u$ (Eq. \ref{Tpro}) for 
the Tinker08 MF at $z=0,1,2,3,4$ (solid black lines). In each panel the virial mass in the MF
is rescaled by the factor shown by the label inside.
The dashed black lines show a cubic extrapolation down to the origin for $z=0$, fitting 
the low limit of the cumulative probability up to the second derivative.
The PS case is displayed in each panel just for visual reference (red dot dashed line, see text)}
\label{cpf}
\end{figure}
 
By using the Tinker08 halo MF in the first equality of Eq. (\ref{Tpro}),  
the cumulative probability function $P(\gamma_u;z)$ can be obtained numerically. 
The result is shown in the top left panel of Fig. \ref{cpf} down to the limit mass at each
epoch (solid black lines for $z=0,1,2,3,4$). 
The corresponding result for the Gaussian (PS) case, with $\dl_c$ rescaled by a factor $0.86$ in order 
to fit the high-mass end of the Tinker08 MF, is also plotted (red dot-dashed curves).
It is interesting to note that for the Tinker08 MF, the dependence of $P(\gamma_u)$ on $z$ is negligible, 
which suggests that $P(\gamma_u)$ as inferred from the simulations deviates only slightly from a universal function.  
In what follows, we will consider that the function $P(\gamma_u)$ is the same at all redshifts, omitting
the explicit argument $z$.

In order to make $P(\gamma_u)$ physically meaningful as a cumulative probability, we need to know its entire 
profile, i.e. down to $\gamma_u= 0$ ($M=0$). 
Then, we proceed first by smoothly extrapolating the Tinker08 $P(\gamma_u)$ applying a cubic polynomial 
calculated with the conditions to be $0$ when $\gamma_u = 0$, and fitting the low limit of the numerical curve 
up to the second derivative. The result is shown by the dashed black curve in the left upper panel of Fig. \ref{cpf}.
The plot highlights the difference between the Tinker08 and the PS cases in the $P(\gamma_u)$ shape when 
$\gamma_u \lesssim 0.5$.
While the latter reaches the origin following a straight line, the first one shows a curved shape. 
The physical reason of such difference can be due to the lack of normalization in the Tinker08 MF;
such a lack of normalization can arise from the presence of diffuse matter in simulations unaccounted 
in the MF (see below).

In the following, we discuss critical issues that appear when a confrontation between analytical approaches 
(e.g., ES and CF) and the N-body simulations is carried out. Then, taking into account these issues, 
we propose an economical strategy that makes compatible for the CF the, the cumulative probability 
$P(\gamma_u$) inferred from simulations and extrapolated to low masses.
In the analytic formalisms, the linear overdense region of mass $M$ is linked to a collapsed halo of the same 
mass (mass conservation; see discussion in \S\S 2.2), and the progenitor distributions at each redshift 
take into account {\it all the available mass as part of collapsed halos}. The question is whether the 
simulations account for the mass in the same way. 

First, in the simulations the range of halo masses is limited; there is no information below the halo mass
resolution limit. If one extrapolates the fitted halo MF to lower masses and integrate the obtained
multiplicity function from 0 to infinity, then one finds a deficit. This implies that the fitted MF 
should change its slope (be steeper) at smaller masses, and/or that some fraction of the mass is actually 
not in the virialized halos.
The latter brings to consideration a second issue: in the simulations a non-negligible fraction of mass is indeed
{\it diffuse} (not in halos). 

The presence of diffuse matter is due to several reasons:
\begin{enumerate}
\item A halo is counted only if it contains tens of particles; groups with less particles are part of what is 
considered as diffuse matter. 

\item Several authors have shown that a significant fraction of the gravitationally bounded particles are 
actually further away from the spherical virial radius (e.g., Prada et al. 2006; Cuesta et al. 2008; Lacerna 
\& Padilla 2011; Anderhalden \& Diemand 2011). 
It is not easy to determine the halo radius that contains all the bounded particles, and likely this radius 
presents a large variation from case to case, depending on environment, epoch, previous assembly history, 
mass, etc. 
Let us call $\psi$ the average mass fraction of matter gravitationally bounded to halos but not accounted in the
conventional spherical virial mass. 

\item Due to true dynamical processes, e.g., when halos collide, some fraction of the particles are ejected from 
the merged system (Wang et al. 2011); let $\varphi$ be the overall fraction of such ejected mass.

The diffuse mass produced by all of these effects may eventually infall on to the growing halos. 
Therefore, at difference of the analytical formalisms, the mass growth of halos in simulations happens also in 
the form of diffuse accretion, which is expected to reduce the minor merger rates as compared to the 
analytical formalisms. In the current simulations, at least 30\% of the $z=0$ halo masses came in diffuse accretion 
(c.f. Genel et al.  2011; Wang et al. 2011).

\item It was suggested also a background of diffuse matter, remnant of the cut-off in the mass 
power spectrum due to the relativistic free-streaming damping (e.g., $\sim 10^{-6}$ \msun\ for the neutralino); at very 
high redshifts, most of this matter due to the cut-off seems to be diffuse (Angulo \& White 2010), but at the redshifts 
of significant halo mass assembly, the effects mentioned above dominate over this.

\end{enumerate}

\begin{figure*} 
\includegraphics[width=120mm, height=120mm, angle=-90]{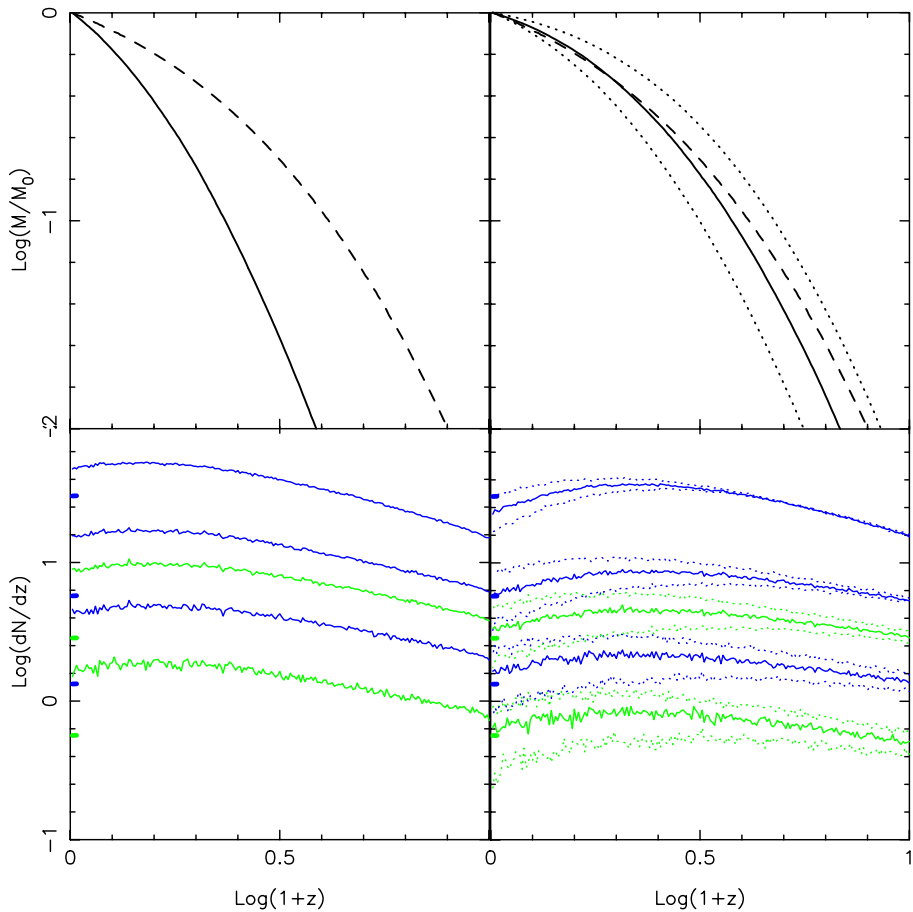}
\caption{{\it  Left panels:} Average MAH (top; solid line) and MRs for different mass-ratio thresholds $\xi$
(bottom; color code as in  Fig. \ref{PSmefr}) for a $M_0 = 10^{13} M_{\odot}$ present-day halo using the 
cumulative probability function derived from the Tinker08 MF by Eq. (\ref{nongauss}). The dashed line 
shows the corresponding mean MAH obtained from the Millennium Simulations (Fakhouri et al. 2010). 
The blue/green thick ticks in the $dN/dz$-axis show the $z=0$ MRs for the different $\xi$ thresholds, as 
reported in Fakhouri et al. (2010; see Fig\ref{PSmefr}). Both the average MAH and MRs do not agree 
with those measured in the Millennium Simulations (see text).
{\it Right panels:}  Same as in left panels but using the cumulative probability function derived from the Tinker08 
MF rescaled in mass by $f= 1.2, 1.3, 1.4$. The solid lines are for $f=1.3$, while the dotted lines for $f=1.2$ and $f=1.4$
(left and right curves in the top panel, respectively, and upper and lower curves for each $\xi$ threshold in the bottom 
panel, respectively). The MAH and MRs agree with those measured in the Millennium Simulation for $f\approx 1.3$ 
(see text).
}
\label{Tmah}
\end{figure*}

The diffuse matter in simulations implies that the mass originally linked to a given overdensity is less than
that defined inside the spherical virial radius. Due to these effects, the conventional virial masses deplete 
on average by the fraction $\varphi+\psi$.  

Now, considering the issues discussed above and that in the CF the diffuse matter is not taken into account, 
we need to make as compatible as possible the cumulative probability $P(\gamma_u$) inferred from simulations 
with the one (correctly normalized from 0 to infinity) to be used in the CF. 
In the case of item (ii) above, the solution is simple: a mass rescaling allows us to recover the mass inside 
the collapsed halo proper of the analytical formalism. With this correction the analytic approach is justified. 
The ejected matter (item iii) may be considered a simple extension of this approximation.
After several experiments, we find that an economical way is using a strategy based on the following ideas:
$a$) a mass rescaling allows to pass from a model (Tinker08) $P(\gamma_u)$ to another $P'(\gamma_u)$,
where the complications of the ejected matter and incorrect halo mass definition in simulations are avoided, 
in such a way that the masses are now roughly compatible with the ones of the analytic case and 
$P'(\gamma_u)$ can be handled with the CF;
$b$) a good empirical criterion is to choose $P'(\gamma_u)$ with a reduced curvature when 
extrapolated to low masses in agreement with the shape of $P_{PS}(\gamma_u)$. 
Finally the choice of the mass rescaling factor is justified by the result, i.e. the MAHs and MRs 
being in agreement with simulations.

Following this strategy, we explore now the effects of a simple constant mass rescaling of the
Tinker08 halo MF on $P'(\gamma_u)$.
The panels of Fig. \ref{cpf} show the cases of virial mass rescaling by factors $f=1.1,~1.2,~1.3,~1.4$, and $1.5$, 
respectively. It is noteworthy that for $f=1.3-1.4$, the extrapolation of $P'(\gamma_u)$ to low $\gamma_u$ 
values is close to a straight line and, in general, this probability function approaches to $P_{PS}(\gamma_u)$.
Moreover, from the papers mentioned above, it can be said that an average halo mass correction by
$\varphi + \psi\sim 30-40\%$ (assumed to be mass independent) is within the uncertainties.
We conclude that this exercise, changing the shape of $P(\gamma_u)$ obtained from simulations so that it 
is closer to the shape of the analytic (PS) Gaussian probability, implies a mass rescaling of $f \approx 1.3$, 
in rough agreement with measures of gravitationally bounded halo mass and diffuse matter considerations in 
simulations.
We do not attempt to follow a more rigorous exercise, for instance introducing a mass dependence in $f$, because
of the large uncertainties involved in the simulation analysis, and because at this point we require only an indicative 
mass correction that encompasses all the complexity implied in passing from the linear overdensity to a virialized structure.  

\subsection{Mass aggregation histories and merger rates according to the mass function from simulations}

The function $P(\gamma_u)$ given by Eq. (\ref{Tpro}) and obtained from the Tinker08 halo MF is the cumulative 
unconditional probability to find an isolated region with overdensity $\dl_c$ and mass $\leq M$ inside the entire
universe, being in this case the {\it cosmic} $\gamma_u = \dl_c / \sqrt{2}\sg_M$.
Now, in order to apply the CF, we need to pass from such unconditional probability to a conditional probability. 
Taking into account the generalized PS Ansatz given by Eq. (\ref{ans}) and according to the definition of $\gamma$ 
in Eq. (\ref{gamma}), the simplest random field compatible with the previous reasoning is 
\begin{equation}
\label{nongauss} 
F(M', \dl' \vert M, \dl ) \equiv \frac{1 - P(\gamma)}{2}.
\end{equation}
In this way, even if the statistics is not any more Gaussian, Eq. (\ref{ans})
by construction is satisfied and the involved mathematics is simple. 
From Fig. \ref{cpf} one can see that our inferred P($\gamma$) from the Tinker08 MFs is different from the one 
corresponding to a Gaussian field (see for a similar conclusion Cole et al. 2008; Neistein et al. 2010). 

The probability $p(M', \dl' \vert M, \dl)~dM'$ to find an isolated region with overdensity $\dl'$ 
and mass between $M'$ and $M' + dM'$ inside an isolated region with overdensity $\dl$ and mass $M$,
according to Eqs. (\ref{psans}) and (\ref{nongauss}), is:
\begin{equation}
\label{probnongauss}
p(M', \dl' \vert M, \dl)~dM' = 
\dfrac{dP(\gamma)}{d \gamma}~\dfrac{d \gamma}{d M'}~dM'.
\end{equation}
We can apply now the Monte Carlo method to find the halo MAH and MRs according to the procedure 
defined in \S\S 2.2.

For $2\ 10^4$ Monte Carlo realizations, we calculate the average MAHs and MRs by using the cumulative 
probability $P(\gamma)$ inferred from the Tinker08 MF (without any mass rescaling, i.e. $f=1$). 
The left panels of Fig. \ref{Tmah} show the average MAH (upper panel, solid line) and
MRs (lower panel, the same five threshold ratios $\xi$ as in Fig. \ref{PSmefr}). 
The corresponding average MAH from the Millennium Simulations (Fakhouri et al. 2010) is 
shown with the dashed line. The predicted MRs are higher at all epochs, but more at lower redshifts, than 
those measured in the simulations by Fakhouri \& Ma (2008) and Fakhouri et al. (2010; thick ticks in the 
$dN/dz$ axis correspond to the $z=0$ MRs as reported by these authors). This produces also a too 
fast mass growth of haloes as seen in the upper panel. 
On the ground of these results, we conclude that the cumulative probability inferred from the original
Tinker08 halo MF leads to halo MAHs and MRs in disagreement with those measured in numerical simulations.

In the right panels of Fig. \ref{Tmah} we present the same as in the left panels
but now for $P(\gamma)$ inferred from the Tinker08 halo MF with the virial mass rescaled by the
factors $f=1.2$, $1.3$, and $1.4$ (see Fig. \ref{cpf}).  The solid curves 
are for $f=1.3$ and the left/right (upper/lower) curves are for  $f=1.2/f=1.4$ in the top (bottom) panel.
The predicted average halo MAH and MRs are now close to those measured in the simulations (certainly within the 
scatter and uncertainties) for the case $f=1.3$.
It is remarkable that the value $f=1.3$ is close to that one which provides a $P'(\gamma_u)$ with a shape 
similar to the shape of $P_{PS}(\gamma_u)$ according to the discussion in \S\S 4.1.
{\it These two findings are completely independent between them. Besides, the value $f\approx 1.3$ seems to be
consistent with measures of the diffuse matter component in the simulations} (see for references \S\S 4.1). 
Note that the MAH and MRs measured from the simulations could change if the virial mass is rescaled; 
however, we do not expect significant changes because in the case of the MAH and MR the physics has to do 
with {\it mass ratios}.

For completeness, we have applied the same procedure to the fitting MF given by Sheth \& Tormen (1999).  
We have obtained practically the same results as for the Tinker08 MF, which for economy we do not repeat here.

Summarizing, through a mass rescaling that makes compatible the masses between simulations and the
analytic framework, we have derived from the halo Tinker08 MF a conditional 
field probability function given by Eq. (\ref{nongauss}) particularly handy in the sense that 
it is compatible with the PS Ansatz (Eq. \ref{ans}), in spite of it deviates from the one of a 
Gaussian density fluctuation field. 
Starting from such probability, our CF allows to calculate the halo MAHs and MRs.
The obtained average MAHs and MRs are consistent with numerical simulations when the virial mass 
in the input Tinker08 halo MF is rescaled.

\begin{figure}
\includegraphics[scale=0.47]{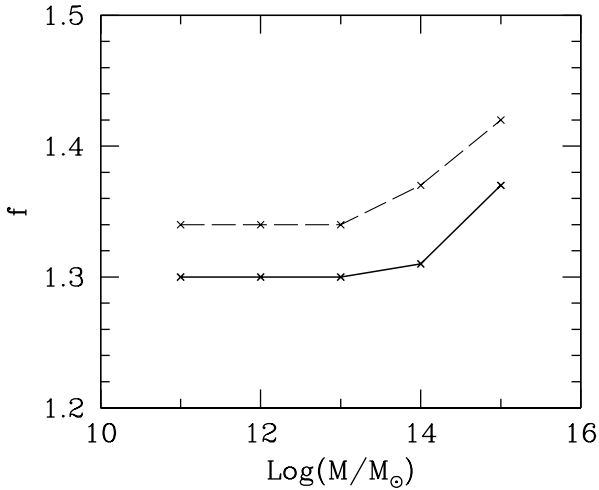} 
\caption{Mass rescaling factor, $f$, required to agree with the average MAH from Fakhouri et al. (2010)
at the redshift one-half (solid line) and one-tenth (dashed line) 
obtained for a wide range of present-day halo masses. }
\label{f-M}
\end{figure}

For economy, we plotted results regarding the average MAH and MRs only for one descendant mass, 
$10^{13}$ \msun. The conclusions are similar for other masses. 
Figure \ref{f-M} shows the value of $f$ required to agree with the Fakhouri et al. (2010) average 
MAHs at the redshift one-half (solid line) and one-tenth (dashed line) of the mass for $z=0$ 
descendants of $10^{11},  10^{12}, 10^{13}, 10^{14},$ and $10^{15}$ \msun. 
As is seen, the mass rescaling factor is roughly the same for masses below $\sim 10^{14}$ \msun, 
something in between 1.30 and 1.34, in order to agree with the MAHs down to one-tenth of the $z=0$ 
descendant mass. For larger masses, this factor should be slightly larger.  
We conclude that, taking into account a mass rescaling of $1.3$, the CF allows for a reasonable 
good description of the halo MFs, and the average MAHs and MRs of halos $\lesssim 10^{14}$ \msun\ 
obtained in numerical simulations.
Probably, a better tuning of this description can be attained by varying $f$ with mass, however, 
this task becomes difficult due to by subtle aspects, for example, the way the averaging is carried out 
when calculating the average MAHs and MRs. We deserve such an exploration for future works.

The level of normalization on the Tinker08 MF influences our result. Suppose that $f_{rn}$ is a 
correction  factor on the MF. The model establishes that $f = 1.3/f_{rn}$, that is a renormalization on 
the MF produces the same effect of a mass rescaling.
If $f_{rn}=1.3$, no mass rescaling should be necessary at all.
A defect in the MF normalization and a defect in the halo mass are related aspects of a same problem.
If the diffuse matter plays the role that we argued in our reasonings above, then the Tinker08 MF does not
include all the dark matter in the cosmological boxes, and consequently it is not normalizable.
Tinker08 indeed acknowledge that this is the case.  

\section{Conclusions and Discussion}

We develop an approach able to connect the overdensities of the density fluctuation field with the 
abundance and assembly history of the virialized haloes. 
Starting from a very general basis, our \textit{conditional formalism} (CF) concerns the inventory of the 
isolated overdense regions given the field conditional probability function of the density fluctuation field. 
This formalism allows us to calculate at the same time and in a simple way: the halo mass function MF 
at any $z$, the halo mass aggregation histories MAHs, and the halo merger rate histories MRs. 
After that, our main goal is to identify a strategy that allows us to use the CF for describing optimally 
the numerical simulation results regarding these three halo statistical and evolutionary features.
In the way of attaining this goal, we arrived to some important conclusions:

\begin{itemize}

\item For the Gaussian density field (the generalized PS Ansatz, Eq. (\ref{ans}), applies), the predicted average halo 
MAH and MRs agree roughly with those measured in current cosmological ($\Lambda$CDM) simulations, 
but the halo MF, as previously found, is overabundant at intermediate masses and deficient at high 
masses as compared to simulations, e.g., with the Tinker08 MF. 

\item The introduction of $\dl_c$ depending on mass (moving barrier; Sheth et al. 2001) instead of 
the constant one, improves the halo MF as compared to simulations, but dramatically spoils
the average MAH and MRs. 
Nonetheless, the major problem with the mass-dependent $\dl_c$ is the excessive sensitivity of 
the MAH and MRs to the redshift step used in the Monte Carlo merger tree construction. 

\item A cumulative field unconditional probability function can be inferred from the halo MF measured 
in simulations (Tinker08), being this almost the same (universal) at different redshifts when expressed
in terms of $\dl_c/\sigma$; its definition has to be complemented by adequately extrapolating it to low masses. 
This inference is particularly suitable regarding the shape when the halo virial mass is rescaled by 
$f\approx 1.3$. 
Remarkably, this rescaling makes the halo mass compatible with the mass used in the analytical formalisms 
having in mind that in simulations not all the mass is in halos. 
On the ground of the CF, the conditional probability function is obtained from the unconditional one by making 
the former compatible with the PS Ansatz, in spite of that the shape of the latter deviates from the one 
corresponding to a Gaussian statistics.
Such a situation allows us to easily use the CF and our Monte Carlo algorithm for calculating the halo MAHs and MRs. 
The obtained average MAHs and MRs depend critically on the mass rescaling factor. 
If this factor, assumed constant, is $f\approx 1.3$, then the agreement with the average MAH
and MRs in simulations becomes remarkable, at least for descendant halos up to $\sim 10^{14}$ \msun.
It is encouraging that the rescaling of the virial mass in the Tinker08 halo MF by a factor 
$f\approx 1.3$ allows both for a well behaved overall cumulative field unconditional probability function
and for halo MAHs and MRs in agreement with simulations. 
It is important to remark that the value of $f$ depends on the resolution limit of numerical simulations.

\item Dark matter in simulations exhibits a rather complex structure.
Not all the mass is counted in the halos defined up to the conventional virial radius, but a diffuse 
component is present because of at least three reasons:
a) the mass resolution limit on halo identification,
b) the not counting of gravitationally bounded particles that are further away the virial radius, and
c) the mass ejection generated by mergers.
The contribution a) has to do with the low mass extension of the MF.
Due to b) and c) the mass accounted in the virialized halo is smaller than in the original overdense
region, producing this a normalization defect in the MF.
The disagreement between the simulation and analytic MFs seems to lie on these last two effects as well as 
on a departure from the Gaussian distribution of the density fluctuations.
The conditional probability function inferred from simulations with our approach and its use in the CF allow 
for an estimate of about $30\%$ for the diffuse mass fraction related to effects b) and c), and show that the 
conditional probability corresponds to a density fluctuation field (slightly) deviated from Gaussianity.
In agreement with the previous argument, the Tinker08 MF does not include all the dark matter 
present in the cosmological boxes and consequently it is not normalizable. 

\end{itemize}

A consistent description of the cosmological simulations results regarding the halo MF, MAHs, and 
MRs through an analytical (statistical) formalism has been achieved. 
The interesting finding is that such a description could not be attained for a Gaussian density 
fluctuation field, as well as without taking into account the complex distribution of dark matter
in simulations. In this sense, our results urge for a revision of two questions: 

\begin{enumerate}

\item {\it The correct accounting in simulations of the mass in virialized halos and of diffuse matter.} 
In the analytical formalisms there is an exact equality between the mass of a linear overdensity and 
that of the collapsed halo used for the counting in the MF, while in the
numerical simulations this is not the case because of effects b) and c) above mentioned.
Therefore, when confronting the halo MF calculated in the analytical formalisms with that one measured in 
simulations, these issues should be taken into account. 
Our approach is a first approximation to this complex problem. Further exploration is necessary. 

\item {\it The correct introduction of the statistics in analytical formalisms.} 
The hierarchical structure formation implies that in the same density field coexist collapsed 
structures at small scales with regions in the linear regime at large scales.  
Then, even if the primordial statistics is Gaussian, at the time the analytical approach is 
applied the density field at the scales of interest could already have deviated from 
Gaussian initial conditions (e.g., Coles \& Jones 1991), as numerical simulations suggest. 

\end{enumerate}

In any case, the CF and the heuristic approach presented here, allow us to overcome some of the 
difficulties related to these questions, and attain a consistent description of the halo MF, MAHs and MRs 
measured in cosmological numerical simulations. 

\section*{Acknowledgments}
We thank the anonymous Referee for his/her comments and suggestions which helped to improve 
this paper.  C. F. is grateful to Dr. Flavio Firmani from the University of Victoria (BC,Ca) for the 
encouraging support to carry out this work. 
V. A. acknowledges CONACyT grant 167332-F for partial funding.

\end{document}